%% file: main.tex
\documentclass[sigconf]{acmart}

\AtBeginDocument{%
  \providecommand\BibTeX{{%
    \normalfont B\kern-0.5em{\scshape i\kern-0.25em b}\kern-0.8em\TeX}}}

\setcopyright{acmcopyright}
\copyrightyear{2021}
\acmYear{2021}
\acmDOI{10.1145/nnnnnnn.nnnnnnn}

\acmBooktitle{Proceedings of ACM SIGIR Workshop on eCommerce (SIGIR eCom’21)}
\acmPrice{15.00}
\acmISBN{978-X-X-XXXX-X/YY/MM}

\input{preamble}

\setcopyright{rightsretained}
\acmConference[SIGIR eCom'21]{ACM SIGIR Workshop on eCommerce}{July 15, 2021}{Virtual Event, Montreal, Canada}
\acmYear{2021}
\copyrightyear{2021}
\makeatletter
\renewcommand\@formatdoi[1]{\ignorespaces}
\makeatother
\acmISBN{}

\begin{document}

\title{Conditional Sequential Slate Optimization}


\author{Yipeng Zhang}
\authornote{Work completed during internship at eBay}
\affiliation{\institution{University of California, Los Angeles}\country{USA}}
\email{zyp5511@g.ucla.edu}

\author{Mingjian Lu}
\affiliation{\institution{eBay Inc}\country{USA}}
\email{mingjlu@ebay.com}

\author{Saratchandra Indrakanti}
\affiliation{\institution{eBay Inc}\country{USA}}
\email{sindrakanti@ebay.com}

\author{Manojkumar Rangasamy Kannadasan}
\authornote{Work completed during working at eBay}
\affiliation{\institution{Facebook Inc}\country{USA}}
\email{mkannadasan@fb.com}

\author{Abraham Bagherjeiran}
\affiliation{\institution{eBay Inc} \country{USA}}
\email{abagherjeiran@ebay.com}

\renewcommand{\shortauthors}{Zhang, et al.}

\begin{abstract}
The top search results matching a user query that are displayed on the first page are critical to the effectiveness and perception of a search system. A search ranking system typically orders the results by independent query-document scores to produce a slate of search results. However, such unilateral scoring methods may fail to capture inter-document dependencies that users are sensitive to, thus producing a sub-optimal slate. Further, in practice, many real-world applications such as e-commerce search require enforcing certain distributional criteria at the slate-level, due to business objectives or long term user retention goals. Unilateral scoring of results does not explicitly support optimizing for such objectives with respect to a slate. Hence, solutions to the slate optimization problem must consider the optimal selection and order of the documents, along with adherence to slate-level distributional criteria. To that end, we propose a hybrid framework extended from traditional slate optimization to solve the conditional slate optimization problem. We introduce conditional sequential slate optimization (CSSO), which jointly learns to optimize for traditional ranking metrics as well as prescribed distribution criteria of documents within the slate. The proposed method can be applied to practical real world problems such as enforcing diversity in e-commerce search results, mitigating bias in top results and personalization of results. Experiments on public datasets and real-world data from e-commerce datasets show that CSSO outperforms popular comparable ranking methods in terms of adherence to distributional criteria while producing comparable or better relevance metrics.
\end{abstract}

\ccsdesc[500]{Computing methodologies~Knowledge representation and reasoning}

\keywords{slate optimization, recommendation system, reinforcement learning, e-commerce}

\maketitle
\section{Introduction}
Ranking search results that match a user query is a critical component of information retrieval (IR) systems that power many popular websites. Typically, learning to rank (LTR) models are developed to score and sort the results, and they are trained based on labels obtained from user engagement with results. While pairwise or listwise loss functions are most commonly employed in training LTR models, scoring the results during inference is usually done in a unilateral fashion. Such an approach of scoring results independently fails to take inter-document dependencies into consideration. However, studies show that the perception of a search result is influenced by other neighboring results \cite{agarwal2015constrained, neighbor_effect, wang2016beyond}. Such behavior is even more pronounced in domains like e-commerce search, where users typically compare and contrast choices presented on a results page as part of their shopping process. Modeling such dependencies is critical within the top results (for instance, results shown on the first page), which users predominantly engage with. Moreover, in many real-world applications, there is a requirement to ensure certain predetermined distributional criteria within results on a search results page. Managing the holistic composition of results within the page is critical towards user perception of a website, and is often driven by business objectives or long terms goals aimed at objectives such as user retention. For instance, ensuring that the spectrum of selection available for a query is represented within the top results is an important objective at eBay. Other examples of such business objectives include showcasing diversity in e-commerce inventory, mitigating bias in top results or personalization of results. Hence, the constitution of an optimal set of top results (documents, or referred to as items in this paper) must take any distributional criteria and inter-result dependencies into account, along with the individual goodness of results. 

 
We refer to the top results for a search query, such as $k$ items shown on the first page as perceived by the user (where $k \ll n$, $n$ is the number of results in the recall), as the slate. In this paper, we present a novel approach and an architecture to generate an optimal slate. An optimal slate is envisioned as a slate that maximizes the likelihood of fulfilling search needs while satisfying predetermined query-specific distributional criteria. We jointly optimize for classical ranking metrics and adherence of the slate to distributional criteria, while modeling intra-slate dependencies. Various approaches have been presented by researchers to produce a set-aware ranking towards slate optimization. Heuristic methods such as Maximal Marginal Relevance (MMR) \cite{MMR}, IA-Select \cite{IA-Select}, and xQuAD \cite{xQuAD} aim to incorporate diversity into a slate or minimize user dissatisfaction as users browse the returned results. These approaches, however, tend to utilize simpler heuristics that may lack the expressive power of the more complex models that are increasingly being used to capture inter-document dependencies. More recent models adopt set-aware ranking architectures \cite{group_wise_ranking,seq2slate}. By modeling a representation of the candidate set of documents and scoring documents cognizant of the candidate set, these approaches aim to produce an optimal slate in a set-aware fashion. For example, groupwise scoring functions (GSF) \cite{group_wise_ranking} proposes multivariate scoring functions to jointly score documents and Seq2Slate \cite{seq2slate} adopts pointer networks to score documents sequentially. While these approaches are set-aware during inference, they are tailored to optimize classic ranking metrics such as  normalized discounted cumulative gain (nDCG). They do not explicitly model the additional distributional criteria, which is an important consideration in practice. To the best of our knowledge, there are no existing neural network based methods directly addressing this specific problem observed in search and specifically e-commerce.

To that end,  we propose conditional sequential slate optimization (CSSO), a novel framework that aims to address the slate optimization problem. It incorporates a sequential decoding approach built upon pointer network, that, during sequential decoding, manages the composition of the slate as per predetermined distributional criteria, while jointly optimizing traditional ranking metrics such as nDCG. This approach treats slate optimization problem as a re-ranking problem, where a \textit{base-ranker} produces a ranking based on a unilateral scoring of items, and the \textit{re-ranker} consumes a set of $n$ ranked items to produce a slate of size $k$, where $k \ll n$. We express query-specific distributional criteria with respect to the slate by means of representative or domain-specific features of the items being ranked. 

The hybrid optimization goal might lead to divergent optimization directions for each term as the model has to balance among ranking quality and all aspects within the predetermined distributional criteria set. Furthermore, due to the existence of non-unique solutions and the non-convex nature of this problem, coupled with the non-differentiable nature of metrics like nDCG, it is hard for traditional gradient based methods to find an optimal slate. To tackle this novel and challenging problem, we employ Reinforcement Learning (RL) for the training of CSSO. RL provides a systematic way to search in the feature space while progressively improving the non-differentiable optimization goal with the aid of policy gradient method. At the same time, with the exploration strategy in RL, the learnt model is able to generalize better than traditional supervised method by efficiently learning the meaningful features from the rollouts.



In the training, we utilize items’ click-through data from IR systems as implicit relevance labels. Within the RL paradigm, a state is represented \textit{jointly} by the current items on the slate, the distance from the predetermined distribution, and the encoded candidate set. An action is tantamount to the item selection process in each step of decoding, and reward is determined on the basis of the slate composition after the decoding process by considering both ranking quality and adherence to distributional criteria. The optimization problem, thus well defined, can be solved by policy gradient in the finite horizon with length $k$, and we adopt a baseline model to reduce the variance in training. 
The key contributions of our work can be summarized as follows:

\begin{itemize}
  \item A framework to jointly optimize a slate for document relevance as well as adherence to a predetermined distribution, thus attempting to address an industry pain point of re-ranking to personalize, mitigate bias or introduce diversity in results.
  \item An enhancement to pointer network to include a conditional structure that facilitates representing slate composition during the decoding phase.
\end{itemize}


In experiments performed on public and proprietary e-commerce datasets, CSSO produces results with comparable ranking metrics and better distribution conformity when compared with state-of-the-art (SOTA) slate optimization methods. On Yahoo \cite{yahoo_dataset} and Web30k \cite{LETOR} datasets with additional simulation for imitating users’ real world behaviors, we demonstrate that both nDCG and GAP (distribution conformity metric which will be defined in the sections that follow) can be simultaneously optimized in the ranking task. Experiments on datasets from eBay.com search sessions show a significant improvement in GAP metrics and  comparable ranking metrics to production ranker. In general, CSSO performs well with respect to both nDCG and GAP. We view CSSO as a practical approach to solve a very relevant industry problem of optimizing a slate for relevance and slate composition based on business objectives. We envision that the approach can be applied to solve problems such as ensuring diversity, mitigating bias, and adhering to business objectives across search and recommendation systems in various web domains including e-commerce, online media and web search.
\section{Related work}
Major LTR methods include pairwise and listwise approaches. Classic works within these two categories are widely used in current ranking production. For pairwise approaches, there are RankSVM \cite{RankSVM} and RankBoost \cite{rankboost}, LambdaMART \cite{LambdaMART}, where the input is a feature vector of each single document and the output is a score for each single document; For listwise approaches such as ListNet \cite{ListNet}, AdaRank \cite{AdaRank} as well as  LightGBM \cite{LightGBM}, the input of these models is a set of document features associated with a query and output is a ranked list. All these approaches focus on learning the optimal way of combining features through training. However, these methods score documents independently and fail to take inter-item dependencies into consideration during inference \cite{indrakanti2019influence}.


During inference, the top-$k$ items in the slate could be recommended in a holistic manner considering them as a set rather than independent items. For example, instead of ranking each item independently, GSF \cite{group_wise_ranking} proposes a deep neural network based algorithm such that the relevance score of a document is determined jointly by multiple documents in the list. Different from just using relevance score to recommend, List-CVAE \cite{List-CVAE} uses a deep generative model to pick k items closest to each desired embedding from the original $n$ items. Moreover some recent works are focusing more on using deep reinforcement learning as an optimization tool and show performant results, such as \cite{rl_rec_1, rl_rec_2, rl_rec_3, rl_rec_4}. The most comparable work to ours is Seq2Slate \cite{seq2slate} which defines the slate optimization problem as a sequential reranking problem. However, these methods mostly focus on the ranking quality of the slate but lack a distributional objective, which is critical in many real world and specifically e-commerce applications.



There are few works directly optimizing for the predetermined distributional criteria on the slate, such as \cite{tandon2020addressing, drosou2010search}. Moreover, diversification algorithms have been used to specifically diversify the search results without explicitly considering the distribution of the ranking results. Heuristic diversification algorithms such as \cite{IA-Select, MMR, xQuAD} serve as a post-processing layer on the ranking score produced by original rankers. Some deep learning algorithms \cite{diverse_loss_1, diverse_loss_2} were also proposed with diversification loss terms to jointly optimize for diversity and ranking quality. However, these works do not account for query-specific distribution preferences, limiting the ability to tailor recommendations to specific queries. 
Our approach aims to expand the scope of query-specific slate optimization to fulfill search needs better and facilitate managing slate composition compared to previous methods.

\section{Approach}
\subsection{Slate optimization overview}

Typically the slate presented to a user is constructed by refining a base ranking result via re-ranking. We define slate optimization as a re-ranking problem that optimizes the slate holistically. For a query $\mathrm{q}$, an ordered candidate set $\mathcal{X}$ consisting of $n$ items is provided by a base ranker. Each item $i$ is associated with a feature vector $\myvec{x}^{(i)}$ ($i=1,\dots,n$) where $\myvec{x}^{(i)}\in \mathbb{R}^m$ and $m$ is the dimension of the feature vector. In our problem, the slate optimization model selects the top-$k$ ($k \ll n$) items from a permutation $\myvec \sigma$ of the ordered candidate set $\mathcal{X}$ such that both the ranking and distribution metrics are jointly optimized. It has shown success in realistic applications in ranking items sequentially, where the re-ranker selects the next best item based on the items already selected. Inspired by the Seq2Seq model \cite{seq2seq}, the re-ranker selects items from the candidate set sequentially, generating $a_t$ with the probability $p(a_t | a_1,\dots, a_{t-1}, \mathcal{X})$, where $a_t \in \{1,\dots,n\}$ denotes the index of the element in the input for the $t^{th}$ selection. The top $k$ ranked elements in the permutation are hereby denoted as $\myvec{\sigma}_k$ i.e. $ \myvec{\sigma}_k = [a_1 , \dots,  a_k ]$.

\subsection{Measure distribution divergence}



%

To rigorously define the parameters of the re-ranker, we first introduce some notations. Within the feature vector of the $i^{th}$ item, $\myvec{x}^{(i)} \in \mathbb{R}^m$, categorical features are one-hot encoded. We denote a categorical variable using an index set $F_j \subset \{1,\dots,m\}$ that stands for indices in the feature vector corresponding to the categorical variable, where $j \in \{1, \dots c\}$ and $c$ is the number of categorical variables. All of the categorical variables of interest can be denoted as $\mathcal{F} = \{F_1, \dots, F_c\}$. 
For example, consider $m=6$, and the feature vector of the $i^{th}$ item $\myvec{x}^{(i)}=[ 0.3,0,1,1,0,5 ]^T$, where the first and sixth features are continuous, and the second, third, fourth and fifth features are one-hot representation of categorical variables $F_1=\{2,3\}$ and $F_2=\{4,5\}$. Here $2,3,4,5$ correspond to the indices of features in item $\myvec{x}^{(i)}$. We define $\myvec{x}^{(i)}_{F_j}$ as the categorical features belonging to $F_j$ for the $i^{th}$ item, i.e. $\myvec{x}^{(i)}_{F_1}=[0,1]^T$, $\myvec{x}^{(i)}_{F_2}=[1,0]^T$.

The slate distribution of a specific categorical variable $F_j$ on the slate of size $k$ is defined as the percentage of each categorical feature belonging to that categorical variable: $\myvec{r}_j = \frac{1}{k} \sum_{t=1}^k \myvec{x}^{(a_t)}_{F_j}$, $j=1,\dots,c$. Note that the sum of each element in $\myvec{r}_j$ equals to $1$. Accordingly, we can now specify the set of predetermined distributional criteria for a query $q$ as $\mathcal{D}_q = \{\myvec{d}_1,\dots,\myvec{d}_c\}$ where $\myvec{d}_j$ is the predetermined distributional criterion as compared with $\myvec{r}_j$. For example, $\myvec{d}_1 = [0.3, 0.7]$ means that the first selected categorical variable is associated with two categories, and we would like the distribution of this feature in the slate, $\myvec{r}_1$, to be: $30\%$ of the first category, $70\%$ of the second category. 
For the generated slate, we would like some of the items' categorical features to follow their corresponding predetermined distributional criteria. In order to measure the distance between a slate distribution $\myvec{r}_j$ and it's distribution criterion $\myvec{d}_j$, we introduce the categorical distribution gap as the metric which is defined as the max of the element-wise absolute distance between $\myvec{r}_j$  and  $\myvec{d}_j$. Furthermore, we define the whole slate distribution gap (GAP) as the average of the categorical distribution gaps
as shown in Equation \ref{eq:GAP}
\begin{equation}  \label{eq:GAP}
    \text{GAP}(\mathcal{D}_q, \{\myvec{r}\}_{j=1}^c)= \frac{1}{c} \sum_{j=1}^c \lVert \myvec{d}_j - \myvec{r}_j\rVert_{\infty}.
\end{equation}

\subsection{Model architecture} 
To build a model that is capable of selecting items sequentially given an input sequence of item features, we adopt the pointer network architecture that can sequentially select items in the input sequence. In general, our CSSO does not only select item based on items already on the slate, but the selection is also based on the conditional information guiding the model to optimize towards the predetermined query-based distributional criteria.

\begin{figure}[H]
	\includegraphics[width=0.45\textwidth]{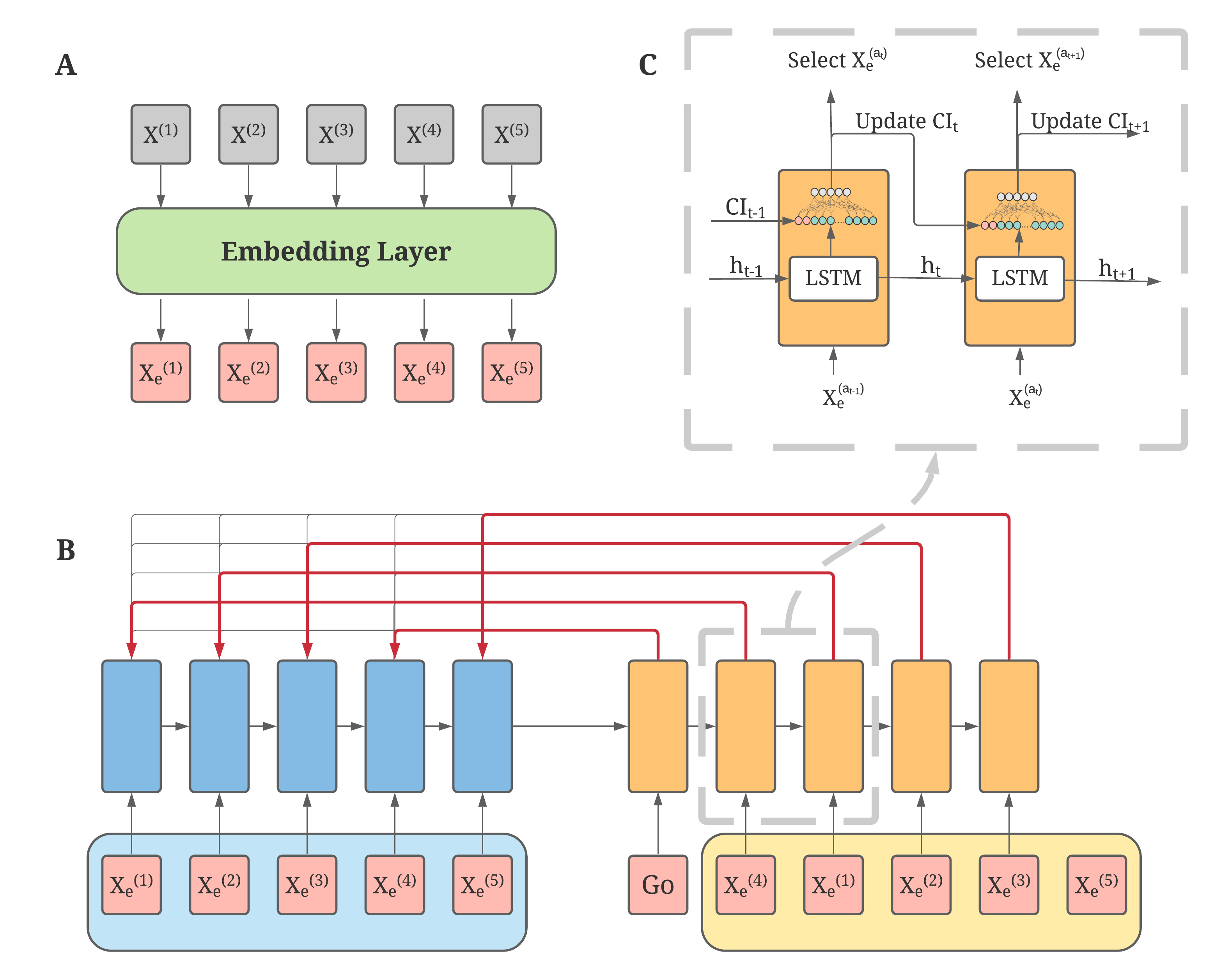}
	\caption{ Architecture of the CSSO model. For the purpose of illustration, we assume that $n = k = 5$ and $x_{F_1}^{(i)}$ ($c = 1$) is a two dimensional vector, and omit the attention mechanism in the figure. (A) The input sequence is transformed by the embedding layer. (B) The embedded input sequence is fed into the encoder LSTM to generate a compact representation of the entire sequence, then the attention decoder selects each item sequentially. (C) In the $t^{th}$ step of decoding, the hidden state of decoder LSTM $h_{t-1}$, the condition information $CI_{t-1}$, and the previous selected item's embedding $\myvec{x}_{e}^{(a_{t-1})}$ are fed into the decoder to guide the model to select the next best item $a_{t}$. Then we compute $CI_{t}$ for next step of decoding.}
	\label{fig:1}
\end{figure}

The typical pointer network consists of encoder and attention decoder, which both contain RNNs to extract sequential information. In our architecture in Figure \ref{fig:1}, we use LSTM \cite{LSTM} for encoder and decoder, while keeping the attention mechanism in the decoder. The items from the input sequence $\myvec{x}^{(1)}, \dots, \myvec{x}^{(n)}$  are first embedded by the embedding layer to generate $\myvec{x}_{e}^{(1)},\dots , \myvec{x}_{e}^{(n)}$. Then the LSTM in the encoder computes a sequence of outputs $H_{en} = (h_1, \dots, h_n)$ iteratively:
\begin{equation}
   h_t = \text{Encoder} \Big(\myvec{x}_{e}^{(t-1)}, h_{t-1} \Big).
\end{equation}
The encoder encodes the embedded input sequence through the hidden state of the LSTM and the last hidden state of the encoder $h_n$ can be viewed as a compact representation of the entire input sequence, which is sent to the decoder as the initial hidden state for the sequential decoding. Meanwhile the output of the encoder is also sent to the decoder for computing the attention matrix. In decoding step $t$, the decoder outputs an $n$-dimensional vector $\myvec{u}_t$ to assign one score for each items in the input sequence. In order to prevent duplicate selection of items in the decoding process, we properly mask $\myvec{u}_t$ before passing it into a softmax to produce a probability distribution, which ensures the probabilities of previously selected items are $0$. Then the index of the next item $a_t$ to put in the slate is sampled following this probability (sampling strategy) or picked greedily (picking the one with maximal probability). After the selection, the new item's embedding $\myvec{x}_{e}^{(a_t)}$ and the output of the hidden state are fed to the decoder for next step decoding. Additionally, in order to guide the model to select items based on the distributional criteria, we feed conditional information $CI_t = \{ \myvec{d}_j - \myvec{r}_j^{(t)}\}_{j=1}^{c} $
where $\myvec{r}^{(t)}_{j} = \frac{1}{t} \sum_{i=1}^t \myvec{x}^{(a_{i})}_{F_j}$
 into the decoder as well, and update $CI_t$ after each decoding step. Therefore, the model is aware of the current status on the slate and selects next items to approach the distributional criteria. The general decoding steps follow the equation below:
\begin{equation}
(h_t, \myvec{u}_t) = \text{Decoder}\Big(h_{t-1},\myvec{x}_e^{(a_{t})}, CI_t\Big).
\end{equation}

\vspace{-0.2cm}

\subsection{Optimal slate}
We define the optimal slate as both satisfying the predetermined distributional criteria and maximizing the ranking metrics. We adopt the GAP metric introduced in previous section to measure the distance of the categorical distributions on the slate to the distributional criteria. For the ranking quality metrics, we use nDCG calculated from the relevance or click-through information $y_i$ for $i^{th}$ item of the query following the reranking order. We introduce a weight $\alpha$ to balance between these two criteria which we want to optimize at the same time, i.e. maximizing the nDCG and minimizing the GAP. Concretely, the final metric of the slate with size $k$ can be written in Equation \ref{eq:reward}.        

\begin{equation}
\label{eq:reward}
\begin{aligned}
\mathcal{R}(\myvec{\sigma}_k, \myvec{y},\mathcal{D}_q,  \{\myvec{r}\}_{j=1}^c) = & \alpha \; (\text{nDCG@k}) \\ 
 & - (1 - \alpha) \text{GAP}(\mathcal{D}_q ,\{\myvec{r}_j\}_{j=1}^c)
\end{aligned}
\end{equation}

Note, the higher $R(\myvec{\sigma}_k, \myvec{y}, \mathcal{D}_q, \{\myvec{r}\}_{j=1}^c )$ represents better permutation $\myvec{\sigma}_k$ in other words, the slate.

\subsection{RL based training} \label{sec:rl}


The employment of RL could enable the model to directly optimize for non-differentiable slate goodness measurement such as nDCG and distribution GAP. The crucial components in reinforcement learning are state, action, and reward.

\newcommand{\p}{p}

The state summarizes the candidate set and the status of the slate generation. The action is to pick an item. An action transits the agent from a state to the next. In the sequential decoding, the agent goes through $k+1$ states, taking $k$ actions. Concretely, in the $t$th step, we include the following components in the state $s_t$: the encoded candidate set $H_{en}$ (for attention mechanism), the condition information $CI_t$, and the selection information till step $t$ embedded in the hidden state $h_t$ of the decoder LSTM.
The action  $a_t \sim \pi_\theta(\cdot|s_t)$ is defined as the agent's selection on the candidate set. 
Recall that $\pi_\theta(\cdot|s_t)$ is the output probability distribution normalized from $\myvec{u}_t$ of the decoder and we have restricted that the probability in previously selected items' locations to be $0$, i.e. $\pi_\theta(a_{t}=a_{i} | s_{t}) = 0, \forall i < t$ to forbid duplicate selections. The optimal policy will be able to generate the optimal slate by taking $k$ actions.



The optimization goal of the CSSO in RL setting is to maximize the expected return $\mathcal{R}(\myvec{\sigma}_k, \myvec{y}, \mathcal{D}_q, \{\myvec{r}\}_{j=1}^c)$ which is a function of permutation $\myvec{\sigma}_k$, item relevance $\myvec{y}$, distributional criteria $\mathcal{D}_q$, and slate distribution  $\{\myvec{r}\}_{j=1}^c$.
\begin{equation}
\underset{\theta}{\text{max}} \quad \mathbb{E}_{a \sim \pi{_\theta}} [\mathcal{R}(\myvec{\sigma}_k, \myvec{y}, \mathcal{D}_q ,\{\myvec{r}\}_{j=1}^c)]
\end{equation}

One can use REINFORCE, a policy gradient method, together with stochastic gradient ascent to optimize policy parameter $\theta$. The gradient is computed by the equation below:
\begin{equation}
\nabla_\theta J(\theta)  = \mathbb{E}_{\pi_\theta} [G_t \nabla_\theta \ln \pi_\theta(a_t \vert s_t)].
\end{equation}
Here the return $G_t$ for $t=1,\dots,k$ is $\mathcal{R}(\myvec{\sigma}_k, \myvec{y},\mathcal{D}_q ,\{\myvec{r}\}_{j=1}^c)$ defined by Equation \ref{eq:reward}, which is a linear combination of nDCG@k and GAP based on the top $k$ items from $\myvec{\sigma}_k$. $G_t$ is the same for all $t$.

By introducing variance reduction technique, the parameters of the model can be updated in a batch with size $N$ via Equation \ref{eq:update_rl}. Note that $\mathcal{R}_i$ is the reward of the $i^{th}$ training sample in the batch:
\begin{equation}
\label{eq:update_rl}
\begin{aligned}
\nabla_\theta J(\theta) & \approx \sum_{i=1}^{N} \frac{1}{N} \left[(\mathcal{R}_i - b)  \sum_{t=1}^{k}\nabla_\theta \ln \pi_\theta(a_t \vert s_t)\right],\\
\text{where} \quad & b  =  \sum_{i=1}^{N} \frac{\mathcal{R}_i}{N}. 
\end{aligned}
\end{equation}

\subsection{Addressing the shortcoming of RL}
The nDCG and GAP used in RL are not differentiable, hence limiting the usage of much more efficient gradient-based methods that directly optimize for the optimal slate. Moreover, RL suffers from low sampling efficiency and is often not stable to train. Therefore we introduce a differentiable objective to evaluate the action at each decoding step to enable direct supervised learning. For the sake of simplicity and consistency, we borrow the notation $s_t$ and $\pi_\theta(\cdot|s_t)$ from section \ref{sec:rl}. $s_t$ is defined as the combination of the encoded candidate set $H_{en}$, the condition $CI_t$, and the selection information till step $t$ embedded in the hidden state $h_t$ of the decoder LSTM, while $\pi_\theta(\cdot|s_t)$ is defined as the output probability distribution normalized from $\myvec{u}_t$ given by the decoder. 

The objective takes two parts into account: the ranking quality and the distributional conformity. For optimizing the ranking quality, we formulate the problem in a supervised learning setting by posing some selection preferences. Instead of optimizing for the whole slate after $k$ steps selections, at each step, the items with higher relevance or click-through labels are preferred among the remaining set. However, there is no unique answer for the decision at each step, so we want the model to equally prefer the items with the same relevance/click-through label. The selection at a single step can therefore be framed as a multiclass classification problem: the model assigns a probability for each of the candidate item. Therefore, at the step $t$ of decoding, we want the model to output a probability distribution $\myvec{p}_t = \pi_\theta(\cdot|s_t)$ aligned with the distribution given by the labels $\myvec{y}_t$ after masking. Specifically, the $p_i$ within vector $\myvec{p}_t$ from previous selections are zero-masked and their corresponding labels are also masked to ensure we only consider the distribution over the labels from the remaining items, i.e.
\begin{equation}
\begin{aligned}
  \myvec{y}_t = [y_1, \dots , y_n] , \quad & \text{where} \quad y_{a_{i}} = 0,\forall i < t.\\
\end{aligned}
\end{equation}

Then we define the step-wise loss for the ranking metric as 
\begin{equation}
 l_{rank}(\myvec{p}_t,\myvec{y}_t) = - \sum_{i}^{n} \frac{y_i}{\sum_j y_j} \log(p_i).\\
\end{equation}

Next, the slate ranking loss can be defined as the weighted sum of each step's loss, where the weights are to mimic the logarithmic decay in nDCG which takes the ranking position into account. At each step, the label $\myvec{y}_t$ and $\myvec{p}_t$ change according to selections from previous steps. Therefore, we define $\mathcal{P} = \{\myvec{p}_t\}_{t=1}^k$ and  $\mathcal{Y} = \{\myvec{y}_t\}_{t=1}^k$ to denote the set of model probabilities and the corresponding labels among $k$ decoding steps respectively.
\begin{equation}
\begin{aligned}
 L_{rank}\left(\mathcal{P},\mathcal{Y}\right) = \sum_{t=1}^{k} w_t\; l_{rank}(\myvec{p}_t,\myvec{y}_t), \; w_t = \dfrac{1}{\log(t + 1)}.
\end{aligned}  
\end{equation}

For optimizing the feature distribution quality, we formulate distribution on the set as the sum of the step-wise distribution selection. At each step, we can compute the step-wise selection distribution,
\begin{equation}
  \myvec{\gamma}^{(t)}_j = \sum_{i=1}^n p_i x^{(i)}_{F_j} ,
\end{equation}
which is the feature distribution among the candidate set marginalizing over the model's selection probabilities. Note that $p_{a_{i}} = 0, \forall i < t$ to prevent double counting and $\myvec{\gamma}^{(t)}_j$ is differentiable, w.r.t. the model parameter $\theta$. Therefore the slate distribution for a specific $j$ is the sum of the step-wise selection distribution.
\begin{equation}
\myvec{r}'_j= \frac{1}{k} \sum_{t=1}^k  \myvec{\gamma}^{(t)}_j \\
\end{equation}
Recall that we define the distance between category distribution on the slate and the distributional criteria as GAP, so we can formulate the differential version of GAP as
\begin{equation}
 \text{GAP}_{\theta}(\mathcal{D}_q, \{\myvec{r}'\}_{j=1}^c)= \frac{1}{c} \sum_{j=1}^c \lVert \myvec{d}_j -  \myvec{r}'_j \rVert_{\infty}.
\end{equation}
By having the slate-wise ranking and distribution quality loss, we again introduce a weight $\alpha$ to balance between the ranking quality and distributional criteria. Additionally we include a hyper-parameter $\beta$ as these two loss terms are not in the same scale.
\begin{equation}
\begin{split}
\mathcal{L}_{\theta}(\{\mathcal{P}, \mathcal{Y}, \{\myvec{r}'\}_{j=1}^c, \mathcal{D}_q) & =\\
\alpha  \beta L_{rank}\left(\mathcal{P}, \mathcal{Y}\right) & + (1-\alpha) \text{GAP}_{\theta}\left(\mathcal{D}_q, \{\myvec{r}'_j\}_{j=1}^c \right)
\end{split}
\end{equation}

The optimization goal of the CSSO in supervised learning setting is to minimize the loss $\mathcal{L}_{\theta}(\mathcal{P}, \mathcal{Y}, \mathcal{D}_q, \{\myvec{r}'\}_{j=1}^c)$ of the slate which is a function of output probability in each step $\mathcal{P} = \{\myvec{p}_t\}_{t=1}^k$, item relevance $\mathcal{Y}$, distributional criteria $\mathcal{D}_q$, and slate distribution $\{\myvec{r}'\}_{j=1}^c$.
\begin{equation}
\min_{\theta} \quad \mathbb{E}_{\pi_\theta} \left[\mathcal{L}_{\theta}(\mathcal{P}, \mathcal{Y}, \mathcal{D}_q, \{\myvec{r}'\}_{j=1}^c)\right]
\end{equation}

Since we need to minimize the loss, we adopt the gradient descent.
The model parameters can be updated similarly as trained by RL via Equation \ref{eq:train_sp}. 

\begin{equation}
\label{eq:train_sp}
\begin{aligned}
\nabla_\theta J(\theta) & \approx \sum_{i=1}^{N} \frac{1}{N} \left[(\mathcal{L}_i - b)  \sum_{t=1}^{k}\nabla_\theta \ln \pi_\theta(a_t \vert s_t) + \mathcal{L}_{{\theta}_i} \right ]\\
\text{where} \quad & b  =  \sum_{i=1}^{N} \frac{\mathcal{L}_i}{N} 
\end{aligned}
\end{equation}

\section{Experiments}
In this section, we measure the performance of our CSSO model and the state-of-the-art on two popular learning-to-rank datasets and one proprietary dataset from ebay.com search session.  
\vspace{-0.5em}
\subsection{Implementation details}
The embedding layer in the policy network (pointer network) is a fully-connected (FC) layer projecting the input features as 256-dimension vectors followed by a ReLU and dropout with a $10\%$ dropout rate. LSTMs in both encoder and decoder have one hidden layer with 256 units. The output hidden state from the decoder is fed into two consecutive FC layers with ReLU activation, $10\%$ dropout, and batch normalization in between. Then softmax is applied to produce the probability distribution with the same length as the candidate sequence. The baseline ($b$) in equation \ref{eq:update_rl} and \ref{eq:train_sp} is the running average with $0.99$ exponential decay to whiten the return. The policy network is trained using the AdaBelief optimizer \cite{adabelief} with the learning rate of $1\mathrm{e}{-4}$ and mini-batches of 1024 training examples.
The training of the policy network uses the sampling strategy. 
At inference time, we report metrics from the greedy decoding as the evaluation results. 
To avoid overfitting, we employ a validation set to validate the model performance.

\subsection{Learning-to-Rank benchmarks}
The experiments are conducted on two popular learning-to-rank datasets which are the Yahoo Learning to Rank Challenge dataset (set 1) and the MSLR-WEB30k (Web30k) dataset.

\subsubsection{Dataset simulation}
There are two barriers preventing the direct usage of the datasets. First, both datasets do not have prescribed query-based distributional criteria. Second, the labels in both public datasets are per-item relevance scores, lacking higher-order interactions between the clicks that are prevalent in real world scenarios. Thus, we augment these two datasets. First, we select binary feature columns with highest in-query variance, i.e. within a query the feature is not highly skewed towards one category. We then establish distributional criteria for each query on selected column feature. Next, we simulate user interactions to incorporate higher-order interactions between the clicks based on methodology adopted from \cite{interaction-procedure}.

The detailed procedure and parameters are in Appendix \ref{app:dataset}. In sum, we simulate $7$ datasets from Yahoo and Web30k datasets with their corresponding query-based distributional criteria objectives as shown in Table \ref{tab:dataset}. In our experiments, we start with optimizing for one distributional criterion using Dataset $\{ 1, 2, 3, 4, 5, 6 \} $ to take a close look at how nDCG and GAP are balanced in the optimal solutions given by various algorithms, and how CSSO generalizes among different datasets. Next, we investigate how CSSO performs over a more complex setup i.e. jointly optimizing two distributional criteria using Dataset 7. 

\vspace{-1em}
\begin{table}[H]
\begin{center}
\caption{Simulated dataset for benchmark}
\resizebox{0.40\textwidth}{!}{
\begin{tabular}{@{}cccc@{}}
\toprule
Number  & DatasetName & BaseRanker & $\mathcal{D}_q$ Column(s) \\ \midrule
1 & Yahoo       & LightGBM   & 628             \\
2 & Yahoo       & MART       & 628             \\
3 & Web30k      & LightGBM   & 95              \\
4 & Web30k      & LightGBM   & 99              \\
5 & Web30k      & LambdaMART & 95              \\
6 & Web30k      & LambdaMART   & 99              \\
7 & Web30k      & LightGBM &  95 $\&$ 99              \\ \bottomrule
\end{tabular}
\label{tab:dataset}
}
\end{center}
\end{table}
\vspace{-1.5em}
\subsubsection{Compare with State-of-The-Art}

We compare CSSO with SOTA ranking models: LightGBM, MART, and LambdaMART using the augmented datasets. As these SOTA methods do not directly optimize for the distributional criteria, we apply a variant of MMR algorithm to compare the model performance fairly. MMR is a reranking layer; by greedily selecting the item of largest score in sequence, it constructs a slate accounting for both distributional criteria and ranking quality. In MMR, the score is calculated by a linear combination of scores from the SOTA ranking models and the $CI_t$ in each step. A factor $\lambda$ is introduced to balance the model's preference between ranking quality and GAP. When $\lambda = 1$, the MMR purely optimizes for nDCG, and when $\lambda = 0$, it prioritizes the GAP over nDCG, and still considers ranking score when candidate items equally satisfy GAP metrics at one step of reranking. We describe our MMR algorithm in detail in Appendix \ref{app:mmr}.




\subsubsection{Experiments}
In our experiments, we set slate length $k = 10$ and all performance metrics are computed on these $10$ items. Since both nDCG and GAP are bounded between $0$ to $1$, we define the slate goodness, $R_s= 0.5 \text{nDCG} - 0.5 \text{GAP} + 0.5$ considering the ranking quality and the distributional criteria evenly. The $R_s$ is one of the main metrics used to compare performances between models, and it is designed to be in range from $0$ to $1$. 

We train two CSSO models with reinforcement learning (\textit{CSSO-RL}) and supervised learning (\textit{CSSO-SL}) respectively. Additionally, we set $\beta = 0.1$ to scale the ranking loss term for CSSO-SL as it is accumulated over the length of the slate. We adopt early stopping on validation set to avoid overfitting and report the $R_s$ on the test set. Since MMR could produce multiple combinations of nDCG and GAP with varying $\lambda$ values, we report the best $R_s$ among all combinations in the following tables. 

\subsubsection*{\textbf{Yahoo dataset}}
\begin{figure}[H]
	\includegraphics[width = 0.45 \textwidth]{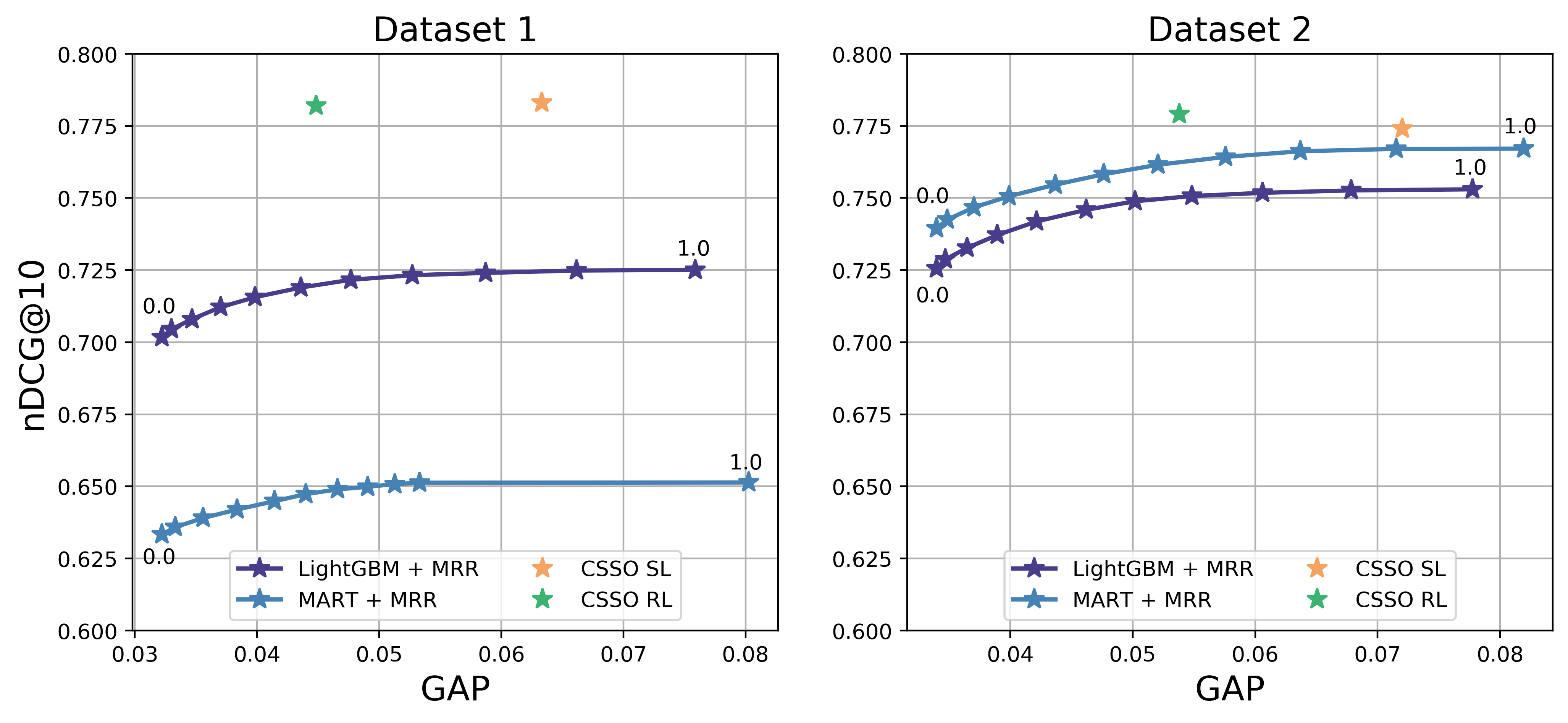}
	\caption{Performance comparisons as we swept the $\lambda$ value in MMR from 0 to 1. Note that a good slate requires a higher nDCG and a lower  GAP. When CSSO-RL and CSSO-SL have similar GAP metrics as SOTA algorithms, our proposed models result in significant improvements in nDCG.}\label{fig:2}
\end{figure}

We generate Dataset $1$ and $2$ based on the Yahoo dataset as described in Table \ref{tab:dataset} and Appendix \ref{app:dataset}.  In Table \ref{tab:yahoo628}, we compare the performance on the test set with LightGMB+MMR (denoted as LightGMB for short) and MART+MMR (MART) with their best $R_s$ respectively. The table displays that on Dataset $1$ and $2$, two CSSO models perform well compared with LightGBM and MART in term of $R_s$. Further, to better visualize the performance comparison with LightGBM and MART using the full range of $\lambda$, we plot the nDCG against GAP for different $\lambda$ in Figure \ref{fig:2}. We observe that with similar GAP metrics, CSSO can have a much better nDCG value than LightGBM and MART.

\vspace{-0.5em}
\begin{table}[H]
\caption{Benchmark on Dataset $1$ and Dataset $2$}
\label{tab:yahoo628}
\resizebox{0.45\textwidth}{!}{
\begin{tabular}{@{}c|ccc|ccc@{}}
\toprule
  & \multicolumn{3}{c|}{Dataset 1} & \multicolumn{3}{c}{Dataset 2} \\ 
Metrics@10              & nDCG $\uparrow$       & GAP $\downarrow$      & $R_s$ $\uparrow$      & nDCG $\uparrow$       & GAP $\downarrow$  & $R_s$ $\uparrow$            \\ \midrule
CSSO-RL                 & 0.782                 & 0.045                 & \textbf{0.869}        & 0.779                 & 0.054             & \textbf{0.863}              \\
CSSO-SL                 & 0.783                 & 0.063                 & \textbf{0.860}        & 0.775                 & 0.072             & 0.852                       \\
LightGBM                & 0.716                 & 0.040                 & 0.838                 & 0.742                 & 0.042             & 0.850                       \\
MART                    & 0.642                 & 0.038                 & 0.802                 & 0.755                 & 0.044             & 0.855                       \\
\bottomrule
\end{tabular}
}
\end{table}

\vspace{-1.3em}

\pagebreak
\subsubsection*{\textbf{Web30k}}
We generate Dataset $3, 4, 5$, and $6$ based on Web30k dataset as described in Table \ref{tab:dataset} and Appendix \ref{app:dataset}. We explore the performance of CSSO to demonstrate the generalizability of the model in different datasets and various distributional criteria. 
Each dataset, in this section, is produced from different feature columns and base rankers. In Table \ref{tab:web30k_3_4}, we compare the performance of CSSO-RL and CSSO-SL versus LightGBM+MMR (LightGBM) and LambdaMART+MMR (LambdaMART) on the data simulated from the same base ranker over different distributional criterion (Dataset $3$ $\&$ Dataset $4$). From Table \ref{tab:web30k_3_4}, we observe that our proposed models significantly outperform the LightGBM and LambdaMART in terms of the $R_s$ which indicates the generated slates with high ranking quality and distributional conformity. 
\begin{table}[h]
	\caption{Benchmark on Dataset $\bm{3}$ $\bm{\&}$ Dataset $\bm{4}$}
	\label{tab:web30k_3_4}
	\resizebox{0.45\textwidth}{!}{
		\begin{tabular}{@{}c|ccc|ccc@{}}
			\toprule
			& \multicolumn{3}{c|}{Dataset 3} & \multicolumn{3}{c}{Dataset 4} \\
			Metrics@10                 & nDCG $\uparrow$ & GAP@95 $\downarrow$ & $R_s$ $\uparrow$  & nDCG $\uparrow$ & GAP@99 $\downarrow$ & $R_s$ $\uparrow$         \\ \midrule
			CSSO-RL             & 0.707           & 0.038            & \textbf{0.834}    & 0.707           & 0.039            & \textbf{0.834}           \\
			CSSO-SL            & 0.704           & 0.077            & \textbf{0.814}    & 0.710           & 0.074            & \textbf{0.818}           \\
			LightGBM              & 0.647           & 0.037            & 0.805             & 0.651           & 0.039            & 0.806                    \\
			LambdaMART            & 0.635           & 0.036            & 0.799             & 0.638           & 0.037            & 0.800                    \\ \bottomrule
		\end{tabular}
	}
\end{table}

Next, we conduct similar experiments as in Table \ref{tab:web30k_3_4} with a different base ranker (Dataset $5$ $\&$ Dataset $6$). We demonstrate the performance comparisons in Table \ref{tab:web30k99}. In this experiment, the CSSO-RL consistently outperforms the SOTA algorithms LightGBM and LambdaMART. However, CSSO-SL does not stand out in term of $R_s$. By comparing the break down of each term in $R_s$, we observe that CSSO-SL still produces a higher nDCG but with worse distributional conformity. Note that the SOTA algorithms can produce extremely low GAP values, which is due to the greedy selection strategy towards the distributional conformity. However, for CSSO-SL, the GAP is parameterized as a function $\text{GAP}_\theta$ then optimized jointly with ranking quality via gradient based method. We hypothesize that in this case, the gradient based method might fail to descend our complex optimization goal due to the non-convex property of the problem, which prevents the CSSO-SL from producing an extremely low GAP value.

\begin{table}[h]
	\caption{\label{tab:web30k99}Benchmark on  Dataset $\bm{5}$ $\bm{\&}$ Dataset $\bm{6}$}
	\label{tab:web30_5_6}
	\resizebox{0.45\textwidth}{!}{
		\begin{tabular}{@{}c|ccc|ccc@{}}
			\toprule
			& \multicolumn{3}{c|}{Dataset 5} & \multicolumn{3}{c}{Dataset 6} \\
			Metrics@10                 & nDCG $\uparrow$ & GAP@95 $\downarrow$ & $R_s$ $\uparrow$  & nDCG $\uparrow$ & GAP@99 $\downarrow$ & $R_s$ $\uparrow$       \\ \midrule
			CSSO-RL             & 0.689           & 0.044            & \textbf{0.823}    & 0.689           & 0.041            & \textbf{0.824}         \\
			CSSO-SL            & 0.679           & 0.081            & 0.799             & 0.683           & 0.078            & 0.803                  \\
			LightGBM                   & 0.660           & 0.038            & 0.811             & 0.660           & 0.041            & 0.809                  \\
			LambdaMART                 & 0.671           & 0.037            & 0.817             & 0.674           & 0.040            & 0.817                  \\ 
			\bottomrule
		\end{tabular}
	}
\end{table}

\begin{figure}[t]
	\includegraphics[width = 0.45 \textwidth]{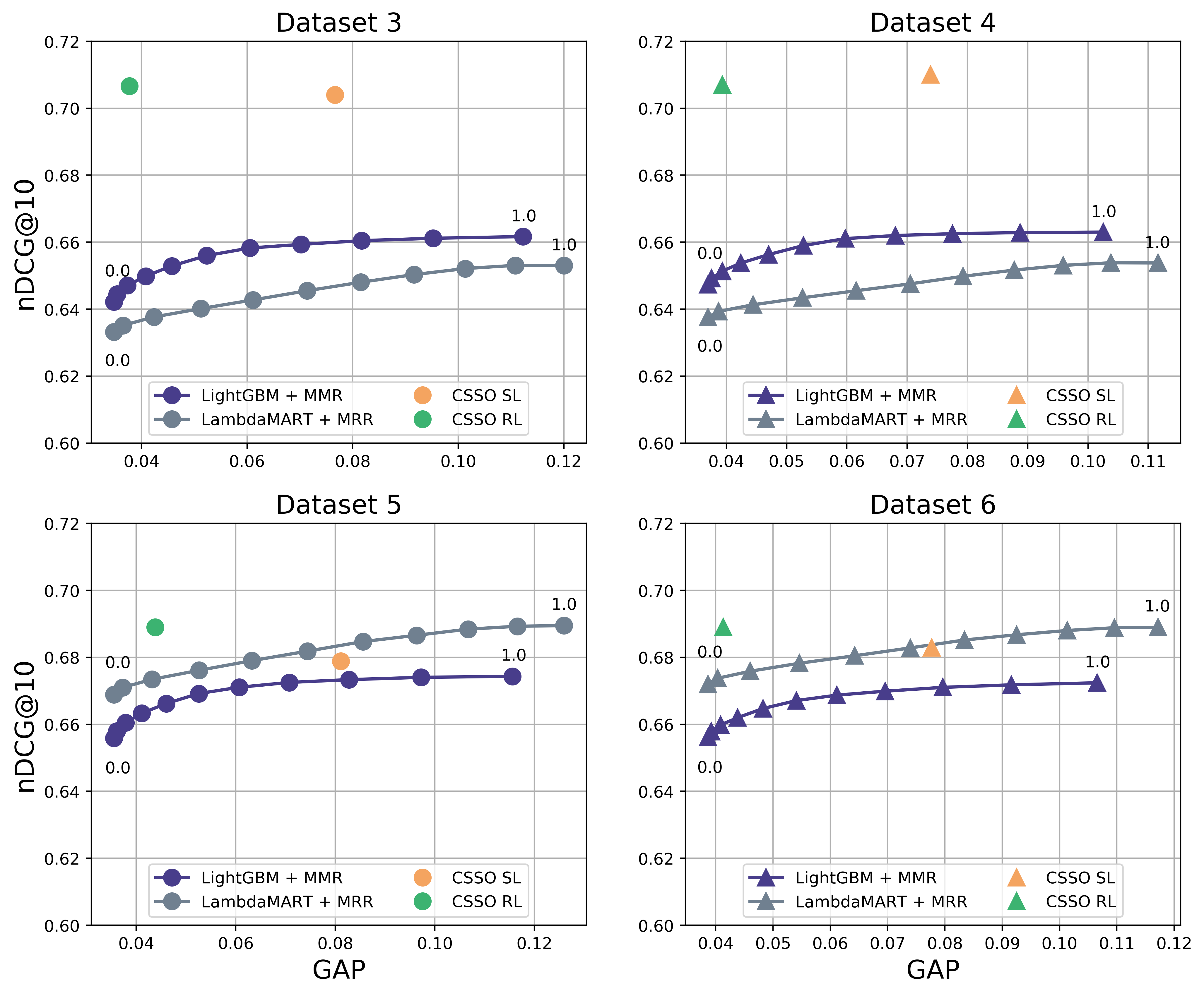}
	\caption{Performance comparisons between CSSO and SOTA algorithms on Dataset $\{3, 4, 5, 6\}$. It demonstrates the CSSO-RL significantly outperforms other SOTA algorithms in different datasets and different distributional criteria.}\label{fig:3}.
\end{figure}

Additionally we plot the trade-off between nDCG and GAP using different $\lambda$ of the LightGBM and LambdaMART, and compare with our proposed framework in Figure \ref{fig:3}. 

In order to demonstrate the performance of CSSO models in a more complex setup, we train them to jointly optimize for two categorical distribution criteria (column 95 $\&$ 99) on Dataset 7. Table \ref{tab:web30kcol2} shows that CSSO models outperform LightGBM and LambdaMART significantly on this complex problem. Compared with optimizing for single distributional criterion, jointly optimizing for multiple distributional criteria is harder due to the potentially contradicting optimization goals, and may result in larger GAP and lower nDCG. Remarkably, both CSSO-SL and CSSO-RL achieve consistent and better performance on all categorical distribution criteria across Table \ref{tab:web30k_3_4} and \ref{tab:web30kcol2}. However, other algorithms have inferior performances. This suggests that CSSO is capable of optimizing for multiple categorical criteria robustly. 


\begin{table}[h]
\caption{Benchmark on Dataset 7}
\label{tab:web30kcol2}
\resizebox{0.45\textwidth}{!}{
\begin{tabular}{@{}c|ccccc@{}}
\toprule
Metrics@10      & nDCG  $\uparrow$      & GAP@95 $\downarrow$   & GAP@99 $\downarrow$   & GAP Avg $\downarrow$  & $R_s$ $\uparrow$      \\ \midrule
CSSO-RL         & 0.704                 & 0.041                 & 0.046                 & 0.044                 & \textbf{0.830}                 \\
CSSO-SL         & 0.705                 & 0.079                 & 0.077                 & 0.078                 & \textbf{0.814}                 \\
LightGBM        &     0.653                  &    0.079                   &   0.074                    &      0.077                 &          0.788             \\
LambdaMART      &   0.649                    &   0.108                    &   0.100                    &    0.104                    &       0.772               \\ 
\bottomrule
\end{tabular}
}
\end{table}

\begin{figure}[h]
	\includegraphics[width=0.45\textwidth]{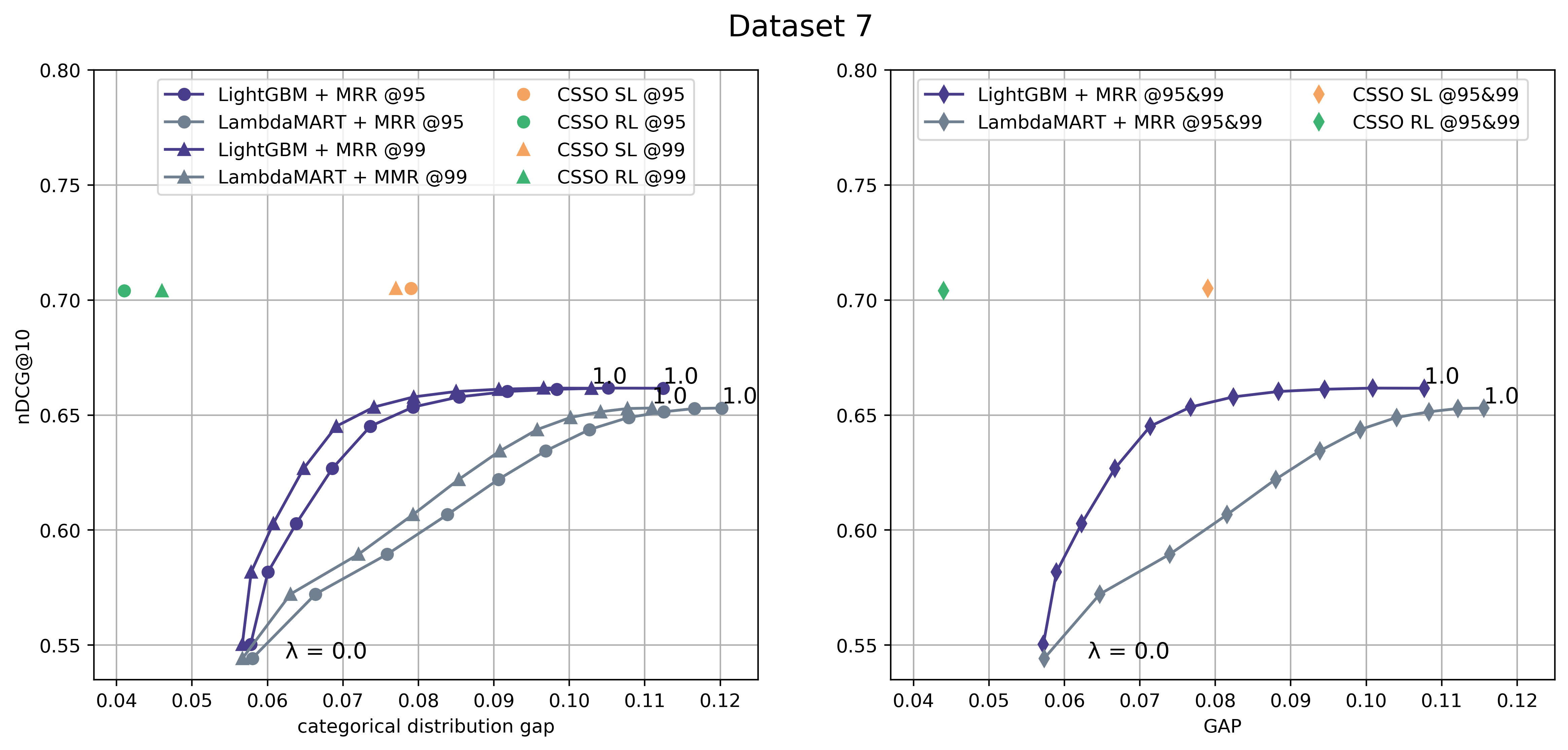}
	\caption{ Performance comparisons for jointly optimizing two distributional criteria: column $95$ $\&$ $99$ on Dataset $7$. We sweep the $\lambda$ for MMR while tracking GAP and nDCG values. The left figure is the detailed gap values for two distributional criteria sharing the same nDCG value, and the right figure is the slate distribution gap (the average of gaps for each distributional criterion). For the purpose of illustration, we merge overlapped annotations of $\lambda$.}\label{fig:4}
\end{figure}

Similarly, we plot the trade-off curve between nDCG and GAP, for these two distributional criteria, in Figure \ref{fig:4}. The left panel is the detailed gap values for each distributional criterion sharing the same nDCG value, and the right panel is GAP (the average of categorical distribution gaps). The superior of both nDCG and GAP value from CSSO-RL and CSSO-SL demonstrates the CSSO models' potential to be versatile and stable tools for slate optimization.

\subsection{Real-world data}
We demonstrate the applicability of the model to a real-world use case using e-commerce search session data from ebay.com. CSSO models are trained as rerankers on top of a base ranker which produces unilateral scoring of eBay items. The aim of reranking, in this case, is to ensure diversity within the search results on the first page. Diversity is quantified based on the composition of the slate with respect to an ideal distribution established for a query computed from historic purchase patterns. 

Training data is sampled from ebay.com search sessions, with user engagement with impressed items such as clicks and purchases used as relevance labels. The training data consists of hundreds of thousands of search queries with the items impressive to users, their corresponding features, and the respective engagement received. Query-specific purchase behaviors are aggregated from historical search sessions to establish a distributional criterion. Desired distributions are computed for each query through an offline process using observed purchase distribution with respect to item-specific categorical features such as condition (for example, new, used, or refurbished conditions). Since labels are based on actual user engagement rather than human-judged samples, inter-item dependencies are inherently captured within the labels. CSSO is trained to optimize for nDCG with respect to the engagement-based relevance labels while adhering to the query-specific distributional criteria. In the reranking process, the input sequence is fed from the production ranker, and the conditional information is based on item-specific categorical features. Offline experiments on eBay search datasets show that CSSO produces a significant reduction in the GAP metric while producing a comparable nDCG with respect to production ranking models as shown in Table \ref{tab:ebay}.

\begin{table}[h]
	\begin{center}
		\caption{\label{tab:ebay}Results from offline experiments on eBay dataset. Metrics are relative to the baseline production ranker and numbers reported as relative change due to confidentiality requirements. The first two rows are the performance of our proposed models for eBay dataset. For the ablation study (the last two rows), we train both models in the same setting but with removal of condition information and name them as CSSO-SL w/o $CI_t$ and CSSO-RL w/o $CI_t$ respectively. The results demonstrate that the utilization of $CI_t$ significantly improves the GAP metrics.}
		\begin{tabular*}{0.35\textwidth}{@{}c|c@{\extracolsep{\fill}}cc@{}}
			\toprule
			Metrics@10          & nDCG   $\uparrow$ & GAP     $\downarrow$  & $R_s$ $\uparrow$      \\ \midrule
			CSSO-SL             & +0.995\%          &    -6.265\%           & \textbf{+2.766}\%              \\
			CSSO-RL             & -0.984\%          &   -7.931\%            & \textbf{+2.687}\%              \\
			CSSO-SL w/o $CI_t$  & +1.065\%          &    -1.942\%           & +1.133\%              \\
			CSSO-RL w/o $CI_t$  & -0.619\%          &    -2.184\%           & +0.613\%              \\ 
			\bottomrule
		\end{tabular*}
	\end{center}
\end{table}

\subsection{Ablation study}

To understand how the conditional information guides the model to optimize for distribution conformity, we surgically remove the $CI_t= \{ \myvec{d}_j - \myvec{r}_j^{(t)}\}_{j=1}^{c}$ in the model while keeping the same training setting; in each decoding step, there are no explicit guidance of the distance to the distribution conformity. We observed the inferior performance of GAP for both \emph{CSSO-SL w/o $CI_t$} and \emph{CSSO-RL w/o $CI_t$} as shown in last two rows of Table \ref{tab:ebay}. While the ranking performance of the models is comparable, the substantially lower GAP metrics of the models with $CI_t$ verifies the significance of introducing conditional information guidance in our CSSO. 

\subsection{\textbf{Observations}}


Experiments show that CSSO-RL usually achieves better GAP than CSSO-SL. Reinforce based method might be able to search with exploration and jump out of some bad local minimum to produce a better slate with high ranking quality and good distribution conformity. However, in-spite of slightly inferior GAP metrics, CSSO-SL provides a good compromise by achieving its best performance at least five times faster than CSSO-RL and even faster than SOTA algorithms, indicating that the gradient-based method does overcome the sampling inefficiency of the reinforce-based method.


\section{Conclusion}

In this paper, we address an industry pain point of producing a slate with predetermined distributional criteria, while jointly optimizing for slate composition and ranking metrics. Conventional practices of employing unilateral scoring of results during inference, or using simple heuristics based rerankers produce sub-optimal slates. More recent approaches of expressive set-aware ranking models optimize exclusively for classic ranking metrics. To that end, we proposed CSSO, a novel reranking architecture building upon pointer network to produce an optimal slate, with optimality defined on a combination of relevance and holistic slate composition. We introduced a conditional structure to the decoding phase to represent a distributional preference and adherence to it. We constructed a reward and loss-function to capture the duality of ranking quality and adherence to desired slate composition and presented approaches to train the model within both the reinforcement learning and supervised learning paradigms. Experiments on Yahoo and Web30k public datasets with simulated click behavior and distributional preferences demonstrate the ability of CSSO to adhere to a prescribed slate composition while improving ranking and relevance metrics. Experiments on proprietary e-commerce datasets using real-world eBay search behavioral data, with the aim of introducing diversity in search results, show the ability of the model to outperform existing production rankers. 

Top search results have a profound impact on user engagement and experience on most IR systems. Web-based businesses have both domain-specific and business-driven priorities that dictate the composition of the top results. We believe that the proposed architecture presents an approach to optimally model such priorities. Several critical real-world problems such as mitigating bias and introducing diversity in top results, addressing under-represented intents, and personalizing a slate based on user-specific priors can be systematically solved using CSSO.

\bibliographystyle{ACM-Reference-Format}
\bibliography{references}

\pagebreak

\appendix
\section*{Appendix}
\section{Dataset simulation procedure}
\label{app:dataset}
In the absence of distributional criteria in the public datasets, we infer $\mathcal{D}_q$ by computing query-wise distributions of the selected categorical variables in the entire item set, i.e. $\myvec{d}_j = \frac{1}{n} \sum_{i=1}^n \myvec{x}^{(i)}_{F_j}$. We treat binary features from the selected dataset as bi-class categorical variables and apply one-hot encoding to them. Specifically, for a query $q$, we infer $\myvec{d}_j$ by computing the percentage of $1$'s and $0$'s in the selected column among all items belonging to it. 

The distributional criteria are trivially satisfied if the feature is query-specific, i.e within a query the feature is highly skewed towards one category. To avoid that scenario, columns are selected because they show the highest within-query variance, allowing for meaningful GAP metric comparison and evaluation.


In order to simulate higher-order interactions between the clicks, we adopt the procedure proposed by Joachims et al .\cite{interaction-procedure} to synthesize click-through data. We first train a base ranker model on the dataset with original relevance labels and rank items within each query on both datasets. Then we adopt the user ``cascade'' model by introducing a parameter $\eta$ in order to quantify the decreasing attention as users browse items from higher rank to lower rank. Specifically, items in each query will have a probability of $1/i^\eta$ to be observed where $i$ is the rank of the item. Subsequently, $\nu$ sequences of users' interactions for a given query are sampled from the observation probability. The labels of sampled items will be converted to click-through labels (clicked: $\{2, 3, 4\}$, skip: $\{0, 1\}$).  In addition, a ``diverse clicks'' mechanism adopted from previous work \cite{seq2slate} is applied such that items similar to previously clicked items will not be clicked. Specifically, we use the median euclidean distance amongst all possible item-pairs for a given query as the threshold to determine if two items are ``similar.'' In the end, we truncate the sequences that are longer than $30$ items to the first $30$ items and pad sequences that are shorter than $30$ items with trailing padding value $v_p$ for all columns. We choose $\nu = 25$ to emulate a real-world scenario in which there are $25$ users interacting with the IR system for each query.

The details for generating the datasets are described below. We first train two different base rankers for each dataset respectively: MART \cite{MART} and LightGBM for the Yahoo dataset, LambdaMART and LightGBM for the Web30k dataset. The base rankers MART and LambdaMART are from the Ranklib package\footnote{\url{https://sourceforge.net/p/lemur/wiki/RankLib/}} and trained with default parameters. The parameter settings of LightGBM are consistent with official LightGBM benchmarks \cite{LightGBM}. In the click simulation, we adopt the values of $\eta$ from previous work \cite{seq2slate}, i.e. $\eta = 0.1$ for Yahoo and $\eta = 0.3$ for Web30k. We first infer the $D_q$ on the original items belonging to each query $q$. 
After click simulation, each query is augmented into $25$ different sub-queries with different subsets of candidates from the original candidate set. All sub-queries share the same query-based distributional criteria $\mathcal{D}_q$. We augment the data with inferred $\mathcal{D}_q$ and simulated click-through labels for selected combination of base ranker and feature selection $\mathcal{F}$ (for specifying $\mathcal{D}_q$). The simulation is applied on both training, validation, and testing set, and the overview of datasets is shown in Table \ref{tab:dataset}.

\section{Variant of MMR}
 
\label{app:mmr}
\begin{algorithm}[tbh]
    \caption{MMR algorithm for one query Q}\label{algo:MMR}
    \textbf{Define}: $C_{F_j}: \mathbb{R}^m \mapsto \mathbb{N}_{+}$, $C_{F_j}(\myvec{x}) := \argmax{(\myvec{x}_{F_j})}$. That is, $C_{F_j}(\myvec{x}^{(i)})$ returns the index of the category of the categorical variable $F_j$\;
    \textbf{Input}: Item categorical variable $F_j$, Item features $\mathcal{X}=\{\myvec{x}^{(i)}\}_{i=1}^n$, Items score from ranker $\myvec{s} \in \mathbb{R}^n$, Coefficient $\lambda$, Distributional criteria $\mathcal{D}_q = \{\myvec{d_j}\}_{j=1}^c$\; 
    \resizebox{0.45\textwidth}{!}{\textbf{Initialize} empty slate $SL$ with length = $k$, 
    $\{\myvec{d_j}'\}_{j=1}^c = \{\myvec{d_j}\}_{j=1}^c$\;} 
    $\myvec{s}'[i] \leftarrow \dfrac{\myvec{s}[i] - \min(\myvec{s})}{\max(\myvec{s}) - \min(\myvec{s})}, \text{ for $i$ from $1$ to $n$ = len$(\myvec{s})$}$\;
    
    \While{$SL$ is not full}{
        $j \leftarrow \displaystyle\argmax_{i \notin SL} \, \Big( \lambda \myvec{s}'[i]+ (1-\lambda) \frac{1}{c} \sum_{j=1}^{c} \myvec{d_j}'[C_{F_j}(\myvec{x}^{(i)})] \Big)$, breaks the tie by larger $\myvec{s}'[i]$ \;
        $\myvec{d}'[C_{F_j}(\myvec{x}^{(j)})] \leftarrow \myvec{d}'[C_{F_j}(\myvec{x}^{(j)})] - \frac{1}{k}$, $j$ = $1$ ,\dots,  $c$\;
        Put $j$ into slate $SL$;
    }
\end{algorithm}

\end{document}

%% file: preamble.tex
\usepackage{algorithmic}
\usepackage{graphicx}
\usepackage{float}
\usepackage{caption}
\usepackage{textcomp}
\usepackage{xcolor}
\usepackage{array}
\usepackage{booktabs}
\usepackage{bm}
\usepackage{lipsum}

\newcommand{\myvec}[1]{\bm{\mathrm{#1}}}
\usepackage[
    ruled, 
    lined,
    longend,
    linesnumbered
  ]{algorithm2e}
\newcommand{\argmax}{\operatorname*{arg\;max}}

%% file: main.bbl

\begin{thebibliography}{31}


\ifx \showCODEN    \undefined \def \showCODEN     #1{\unskip}     \fi
\ifx \showDOI      \undefined \def \showDOI       #1{#1}\fi
\ifx \showISBNx    \undefined \def \showISBNx     #1{\unskip}     \fi
\ifx \showISBNxiii \undefined \def \showISBNxiii  #1{\unskip}     \fi
\ifx \showISSN     \undefined \def \showISSN      #1{\unskip}     \fi
\ifx \showLCCN     \undefined \def \showLCCN      #1{\unskip}     \fi
\ifx \shownote     \undefined \def \shownote      #1{#1}          \fi
\ifx \showarticletitle \undefined \def \showarticletitle #1{#1}   \fi
\ifx \showURL      \undefined \def \showURL       {\relax}        \fi
\providecommand\bibfield[2]{#2}
\providecommand\bibinfo[2]{#2}
\providecommand\natexlab[1]{#1}
\providecommand\showeprint[2][]{arXiv:#2}

\bibitem[\protect\citeauthoryear{Agarwal, Chatterjee, Yang, and Zhang}{Agarwal
  et~al\mbox{.}}{2015}]%
        {agarwal2015constrained}
\bibfield{author}{\bibinfo{person}{Deepak Agarwal}, \bibinfo{person}{Shaunak
  Chatterjee}, \bibinfo{person}{Yang Yang}, {and} \bibinfo{person}{Liang
  Zhang}.} \bibinfo{year}{2015}\natexlab{}.
\newblock \showarticletitle{Constrained optimization for homepage relevance}.
  In \bibinfo{booktitle}{\emph{Proceedings of the 24th International Conference
  on World Wide Web}}. \bibinfo{pages}{375--384}.
\newblock


\bibitem[\protect\citeauthoryear{Agrawal, Gollapudi, Halverson, and
  Ieong}{Agrawal et~al\mbox{.}}{2009}]%
        {IA-Select}
\bibfield{author}{\bibinfo{person}{Rakesh Agrawal}, \bibinfo{person}{Sreenivas
  Gollapudi}, \bibinfo{person}{Alan Halverson}, {and} \bibinfo{person}{Samuel
  Ieong}.} \bibinfo{year}{2009}\natexlab{}.
\newblock \showarticletitle{Diversifying Search Results}. In
  \bibinfo{booktitle}{\emph{International Conference on Web Search and Data
  Mining (WSDM)} (\bibinfo{edition}{international conference on web search and
  data mining (wsdm)} ed.)}. \bibinfo{publisher}{Association for Computing
  Machinery, Inc.}
\newblock
\urldef\tempurl%
\url{https://www.microsoft.com/en-us/research/publication/diversifying-search-results/}
\showURL{%
\tempurl}


\bibitem[\protect\citeauthoryear{Ai, Wang, Bruch, Golbandi, Bendersky, and
  Najork}{Ai et~al\mbox{.}}{2019}]%
        {group_wise_ranking}
\bibfield{author}{\bibinfo{person}{Qingyao Ai}, \bibinfo{person}{Xuanhui Wang},
  \bibinfo{person}{Sebastian Bruch}, \bibinfo{person}{Nadav Golbandi},
  \bibinfo{person}{Michael Bendersky}, {and} \bibinfo{person}{Marc Najork}.}
  \bibinfo{year}{2019}\natexlab{}.
\newblock \showarticletitle{Learning groupwise multivariate scoring functions
  using deep neural networks}. In \bibinfo{booktitle}{\emph{Proceedings of the
  2019 ACM SIGIR International Conference on Theory of Information Retrieval}}.
  \bibinfo{pages}{85--92}.
\newblock


\bibitem[\protect\citeauthoryear{Bai, Guan, and Wang}{Bai
  et~al\mbox{.}}{2019}]%
        {rl_rec_1}
\bibfield{author}{\bibinfo{person}{Xueying Bai}, \bibinfo{person}{Jian Guan},
  {and} \bibinfo{person}{Hongning Wang}.} \bibinfo{year}{2019}\natexlab{}.
\newblock \showarticletitle{A Model-Based Reinforcement Learning with
  Adversarial Training for Online Recommendation}. In
  \bibinfo{booktitle}{\emph{Advances in Neural Information Processing Systems
  32: Annual Conference on Neural Information Processing Systems 2019, NeurIPS
  2019, December 8-14, 2019, Vancouver, BC, Canada}},
  \bibfield{editor}{\bibinfo{person}{Hanna~M. Wallach}, \bibinfo{person}{Hugo
  Larochelle}, \bibinfo{person}{Alina Beygelzimer}, \bibinfo{person}{Florence
  d'Alch{\'{e}}{-}Buc}, \bibinfo{person}{Emily~B. Fox}, {and}
  \bibinfo{person}{Roman Garnett}} (Eds.). \bibinfo{pages}{10734--10745}.
\newblock
\urldef\tempurl%
\url{https://proceedings.neurips.cc/paper/2019/hash/e49eb6523da9e1c347bc148ea8ac55d3-Abstract.html}
\showURL{%
\tempurl}


\bibitem[\protect\citeauthoryear{Bello, Kulkarni, Jain, Boutilier, Chi, Eban,
  Luo, Mackey, and Meshi}{Bello et~al\mbox{.}}{2019}]%
        {seq2slate}
\bibfield{author}{\bibinfo{person}{Irwan Bello}, \bibinfo{person}{Sayali
  Kulkarni}, \bibinfo{person}{Sagar Jain}, \bibinfo{person}{Craig Boutilier},
  \bibinfo{person}{Ed Chi}, \bibinfo{person}{Elad Eban},
  \bibinfo{person}{Xiyang Luo}, \bibinfo{person}{Alan Mackey}, {and}
  \bibinfo{person}{Ofer Meshi}.} \bibinfo{year}{2019}\natexlab{}.
\newblock \bibinfo{booktitle}{\emph{Seq2Slate: Re-ranking and Slate
  Optimization with RNNs}}.
\newblock \bibinfo{type}{{T}echnical {R}eport}.
\newblock
\urldef\tempurl%
\url{https://arxiv.org/abs/1810.02019}
\showURL{%
\tempurl}


\bibitem[\protect\citeauthoryear{Cao, Qin, Liu, Tsai, and Li}{Cao
  et~al\mbox{.}}{2007}]%
        {ListNet}
\bibfield{author}{\bibinfo{person}{Zhe Cao}, \bibinfo{person}{Tao Qin},
  \bibinfo{person}{Tie-Yan Liu}, \bibinfo{person}{Ming-Feng Tsai}, {and}
  \bibinfo{person}{Hang Li}.} \bibinfo{year}{2007}\natexlab{}.
\newblock \showarticletitle{Learning to Rank: From Pairwise Approach to
  Listwise Approach}. In \bibinfo{booktitle}{\emph{Proceedings of the 24th
  International Conference on Machine Learning}} (Corvalis, Oregon, USA)
  \emph{(\bibinfo{series}{ICML '07})}. \bibinfo{publisher}{Association for
  Computing Machinery}, \bibinfo{address}{New York, NY, USA},
  \bibinfo{pages}{129–136}.
\newblock
\showISBNx{9781595937933}
\urldef\tempurl%
\url{https://doi.org/10.1145/1273496.1273513}
\showDOI{\tempurl}


\bibitem[\protect\citeauthoryear{Carbonell and Goldstein}{Carbonell and
  Goldstein}{1998}]%
        {MMR}
\bibfield{author}{\bibinfo{person}{Jaime Carbonell} {and} \bibinfo{person}{Jade
  Goldstein}.} \bibinfo{year}{1998}\natexlab{}.
\newblock \showarticletitle{The use of MMR, diversity-based reranking for
  reordering documents and producing summaries}. In
  \bibinfo{booktitle}{\emph{Proceedings of the 21st annual international ACM
  SIGIR conference on Research and development in information retrieval}}.
  \bibinfo{pages}{335--336}.
\newblock


\bibitem[\protect\citeauthoryear{Chapelle and Chang}{Chapelle and
  Chang}{2010}]%
        {yahoo_dataset}
\bibfield{author}{\bibinfo{person}{Olivier Chapelle} {and} \bibinfo{person}{Yi
  Chang}.} \bibinfo{year}{2010}\natexlab{}.
\newblock \showarticletitle{Yahoo! Learning to Rank Challenge Overview}. In
  \bibinfo{booktitle}{\emph{Proceedings of the 2010 International Conference on
  Yahoo! Learning to Rank Challenge - Volume 14}} (Haifa, Israel)
  \emph{(\bibinfo{series}{YLRC'10})}. \bibinfo{publisher}{JMLR.org},
  \bibinfo{pages}{1–24}.
\newblock


\bibitem[\protect\citeauthoryear{Chen, Dai, Cai, Zhang, Wang, Tang, Zhang, and
  Yu}{Chen et~al\mbox{.}}{2019b}]%
        {rl_rec_2}
\bibfield{author}{\bibinfo{person}{Haokun Chen}, \bibinfo{person}{Xinyi Dai},
  \bibinfo{person}{Han Cai}, \bibinfo{person}{Weinan Zhang},
  \bibinfo{person}{Xuejian Wang}, \bibinfo{person}{Ruiming Tang},
  \bibinfo{person}{Yuzhou Zhang}, {and} \bibinfo{person}{Yong Yu}.}
  \bibinfo{year}{2019}\natexlab{b}.
\newblock \showarticletitle{Large-scale interactive recommendation with
  tree-structured policy gradient}. In \bibinfo{booktitle}{\emph{Proceedings of
  the AAAI Conference on Artificial Intelligence}}, Vol.~\bibinfo{volume}{33}.
  \bibinfo{pages}{3312--3320}.
\newblock


\bibitem[\protect\citeauthoryear{Chen, Beutel, Covington, Jain, Belletti, and
  Chi}{Chen et~al\mbox{.}}{2019a}]%
        {rl_rec_3}
\bibfield{author}{\bibinfo{person}{Minmin Chen}, \bibinfo{person}{Alex Beutel},
  \bibinfo{person}{Paul Covington}, \bibinfo{person}{Sagar Jain},
  \bibinfo{person}{Francois Belletti}, {and} \bibinfo{person}{Ed~H Chi}.}
  \bibinfo{year}{2019}\natexlab{a}.
\newblock \showarticletitle{Top-k off-policy correction for a REINFORCE
  recommender system}. In \bibinfo{booktitle}{\emph{Proceedings of the Twelfth
  ACM International Conference on Web Search and Data Mining}}.
  \bibinfo{pages}{456--464}.
\newblock


\bibitem[\protect\citeauthoryear{Drosou and Pitoura}{Drosou and
  Pitoura}{2010}]%
        {drosou2010search}
\bibfield{author}{\bibinfo{person}{Marina Drosou} {and}
  \bibinfo{person}{Evaggelia Pitoura}.} \bibinfo{year}{2010}\natexlab{}.
\newblock \showarticletitle{Search result diversification}.
\newblock \bibinfo{journal}{\emph{ACM SIGMOD Record}} \bibinfo{volume}{39},
  \bibinfo{number}{1} (\bibinfo{year}{2010}), \bibinfo{pages}{41--47}.
\newblock


\bibitem[\protect\citeauthoryear{Freund, Iyer, Schapire, and Singer}{Freund
  et~al\mbox{.}}{2003}]%
        {rankboost}
\bibfield{author}{\bibinfo{person}{Yoav Freund}, \bibinfo{person}{Raj Iyer},
  \bibinfo{person}{Robert~E Schapire}, {and} \bibinfo{person}{Yoram Singer}.}
  \bibinfo{year}{2003}\natexlab{}.
\newblock \showarticletitle{An efficient boosting algorithm for combining
  preferences}.
\newblock \bibinfo{journal}{\emph{Journal of machine learning research}}
  \bibinfo{volume}{4}, \bibinfo{number}{Nov} (\bibinfo{year}{2003}),
  \bibinfo{pages}{933--969}.
\newblock


\bibitem[\protect\citeauthoryear{Friedman}{Friedman}{2001}]%
        {MART}
\bibfield{author}{\bibinfo{person}{Jerome Friedman}.}
  \bibinfo{year}{2001}\natexlab{}.
\newblock \showarticletitle{Greedy Function Approximation: A Gradient Boosting
  Machine}.
\newblock \bibinfo{journal}{\emph{Annals of Statistics}}  \bibinfo{volume}{29}
  (\bibinfo{date}{10} \bibinfo{year}{2001}), \bibinfo{pages}{1189--1232}.
\newblock
\urldef\tempurl%
\url{https://doi.org/10.2307/2699986}
\showDOI{\tempurl}


\bibitem[\protect\citeauthoryear{Hochreiter and Schmidhuber}{Hochreiter and
  Schmidhuber}{1997}]%
        {LSTM}
\bibfield{author}{\bibinfo{person}{Sepp Hochreiter} {and}
  \bibinfo{person}{J{\"u}rgen Schmidhuber}.} \bibinfo{year}{1997}\natexlab{}.
\newblock \showarticletitle{Long short-term memory}.
\newblock \bibinfo{journal}{\emph{Neural computation}} \bibinfo{volume}{9},
  \bibinfo{number}{8} (\bibinfo{year}{1997}), \bibinfo{pages}{1735--1780}.
\newblock


\bibitem[\protect\citeauthoryear{Hu, Da, Zeng, Yu, and Xu}{Hu
  et~al\mbox{.}}{2018}]%
        {rl_rec_4}
\bibfield{author}{\bibinfo{person}{Yujing Hu}, \bibinfo{person}{Qing Da},
  \bibinfo{person}{Anxiang Zeng}, \bibinfo{person}{Yang Yu}, {and}
  \bibinfo{person}{Yinghui Xu}.} \bibinfo{year}{2018}\natexlab{}.
\newblock \showarticletitle{Reinforcement learning to rank in e-commerce search
  engine: Formalization, analysis, and application}. In
  \bibinfo{booktitle}{\emph{Proceedings of the 24th ACM SIGKDD International
  Conference on Knowledge Discovery \& Data Mining}}.
  \bibinfo{pages}{368--377}.
\newblock


\bibitem[\protect\citeauthoryear{Indrakanti, Strunjas, Tandon, and
  Kannadasan}{Indrakanti et~al\mbox{.}}{2019}]%
        {indrakanti2019influence}
\bibfield{author}{\bibinfo{person}{Saratchandra Indrakanti},
  \bibinfo{person}{Svetlana Strunjas}, \bibinfo{person}{Shubhangi Tandon},
  {and} \bibinfo{person}{Manojkumar Kannadasan}.}
  \bibinfo{year}{2019}\natexlab{}.
\newblock \showarticletitle{Influence of Neighborhood on the Preference of an
  Item in eCommerce Search}. In \bibinfo{booktitle}{\emph{2019 IEEE
  International Conference on Big Data (Big Data)}}. IEEE,
  \bibinfo{pages}{2284--2288}.
\newblock


\bibitem[\protect\citeauthoryear{Jiang, Gowal, Qian, Mann, and Rezende}{Jiang
  et~al\mbox{.}}{2019}]%
        {List-CVAE}
\bibfield{author}{\bibinfo{person}{Ray Jiang}, \bibinfo{person}{Sven Gowal},
  \bibinfo{person}{Yuqiu Qian}, \bibinfo{person}{Timothy Mann}, {and}
  \bibinfo{person}{Danilo~J. Rezende}.} \bibinfo{year}{2019}\natexlab{}.
\newblock \showarticletitle{Beyond Greedy Ranking: Slate Optimization via
  List-{CVAE}}. In \bibinfo{booktitle}{\emph{International Conference on
  Learning Representations}}.
\newblock
\urldef\tempurl%
\url{https://openreview.net/forum?id=r1xX42R5Fm}
\showURL{%
\tempurl}


\bibitem[\protect\citeauthoryear{Jiang, Wen, Dou, Zhao, Nie, and Yue}{Jiang
  et~al\mbox{.}}{2017}]%
        {diverse_loss_1}
\bibfield{author}{\bibinfo{person}{Zhengbao Jiang}, \bibinfo{person}{Ji-Rong
  Wen}, \bibinfo{person}{Zhicheng Dou}, \bibinfo{person}{Wayne~Xin Zhao},
  \bibinfo{person}{Jian-Yun Nie}, {and} \bibinfo{person}{Ming Yue}.}
  \bibinfo{year}{2017}\natexlab{}.
\newblock \showarticletitle{Learning to diversify search results via subtopic
  attention}. In \bibinfo{booktitle}{\emph{Proceedings of the 40th
  international ACM SIGIR Conference on Research and Development in Information
  Retrieval}}. \bibinfo{pages}{545--554}.
\newblock


\bibitem[\protect\citeauthoryear{Joachims}{Joachims}{2002}]%
        {RankSVM}
\bibfield{author}{\bibinfo{person}{Thorsten Joachims}.}
  \bibinfo{year}{2002}\natexlab{}.
\newblock \showarticletitle{Optimizing Search Engines Using Clickthrough Data}.
  In \bibinfo{booktitle}{\emph{Proceedings of the Eighth ACM SIGKDD
  International Conference on Knowledge Discovery and Data Mining}} (Edmonton,
  Alberta, Canada) \emph{(\bibinfo{series}{KDD '02})}.
  \bibinfo{publisher}{Association for Computing Machinery},
  \bibinfo{address}{New York, NY, USA}, \bibinfo{pages}{133–142}.
\newblock
\showISBNx{158113567X}
\urldef\tempurl%
\url{https://doi.org/10.1145/775047.775067}
\showDOI{\tempurl}


\bibitem[\protect\citeauthoryear{Joachims, Granka, Pan, Hembrooke, and
  Gay}{Joachims et~al\mbox{.}}{2005}]%
        {neighbor_effect}
\bibfield{author}{\bibinfo{person}{Thorsten Joachims}, \bibinfo{person}{Laura
  Granka}, \bibinfo{person}{Bing Pan}, \bibinfo{person}{Helene Hembrooke},
  {and} \bibinfo{person}{Geri Gay}.} \bibinfo{year}{2005}\natexlab{}.
\newblock \showarticletitle{Accurately Interpreting Clickthrough Data as
  Implicit Feedback}. In \bibinfo{booktitle}{\emph{Proceedings of the 28th
  Annual International ACM SIGIR Conference on Research and Development in
  Information Retrieval}} (Salvador, Brazil) \emph{(\bibinfo{series}{SIGIR
  '05})}. \bibinfo{publisher}{Association for Computing Machinery},
  \bibinfo{address}{New York, NY, USA}, \bibinfo{pages}{154–161}.
\newblock
\showISBNx{1595930345}
\urldef\tempurl%
\url{https://doi.org/10.1145/1076034.1076063}
\showDOI{\tempurl}


\bibitem[\protect\citeauthoryear{Joachims, Swaminathan, and Schnabel}{Joachims
  et~al\mbox{.}}{2017}]%
        {interaction-procedure}
\bibfield{author}{\bibinfo{person}{Thorsten Joachims}, \bibinfo{person}{Adith
  Swaminathan}, {and} \bibinfo{person}{Tobias Schnabel}.}
  \bibinfo{year}{2017}\natexlab{}.
\newblock \showarticletitle{Unbiased learning-to-rank with biased feedback}. In
  \bibinfo{booktitle}{\emph{Proceedings of the Tenth ACM International
  Conference on Web Search and Data Mining}}. \bibinfo{pages}{781--789}.
\newblock


\bibitem[\protect\citeauthoryear{Ke, Meng, Finley, Wang, Chen, Ma, Ye, and
  Liu}{Ke et~al\mbox{.}}{2017}]%
        {LightGBM}
\bibfield{author}{\bibinfo{person}{Guolin Ke}, \bibinfo{person}{Qi Meng},
  \bibinfo{person}{Thomas Finley}, \bibinfo{person}{Taifeng Wang},
  \bibinfo{person}{Wei Chen}, \bibinfo{person}{Weidong Ma},
  \bibinfo{person}{Qiwei Ye}, {and} \bibinfo{person}{Tie-Yan Liu}.}
  \bibinfo{year}{2017}\natexlab{}.
\newblock \showarticletitle{LightGBM: A Highly Efficient Gradient Boosting
  Decision Tree}. In \bibinfo{booktitle}{\emph{Advances in Neural Information
  Processing Systems}}, \bibfield{editor}{\bibinfo{person}{I.~Guyon},
  \bibinfo{person}{U.~V. Luxburg}, \bibinfo{person}{S.~Bengio},
  \bibinfo{person}{H.~Wallach}, \bibinfo{person}{R.~Fergus},
  \bibinfo{person}{S.~Vishwanathan}, {and} \bibinfo{person}{R.~Garnett}}
  (Eds.), Vol.~\bibinfo{volume}{30}. \bibinfo{publisher}{Curran Associates,
  Inc.}, \bibinfo{pages}{3146--3154}.
\newblock
\urldef\tempurl%
\url{https://proceedings.neurips.cc/paper/2017/file/6449f44a102fde848669bdd9eb6b76fa-Paper.pdf}
\showURL{%
\tempurl}


\bibitem[\protect\citeauthoryear{Qin and Liu}{Qin and Liu}{2013}]%
        {LETOR}
\bibfield{author}{\bibinfo{person}{Tao Qin} {and} \bibinfo{person}{Tie{-}Yan
  Liu}.} \bibinfo{year}{2013}\natexlab{}.
\newblock \showarticletitle{Introducing {LETOR} 4.0 Datasets}.
\newblock \bibinfo{journal}{\emph{CoRR}}  \bibinfo{volume}{abs/1306.2597}
  (\bibinfo{year}{2013}).
\newblock
\urldef\tempurl%
\url{http://arxiv.org/abs/1306.2597}
\showURL{%
\tempurl}


\bibitem[\protect\citeauthoryear{Santos, Macdonald, and Ounis}{Santos
  et~al\mbox{.}}{2010}]%
        {xQuAD}
\bibfield{author}{\bibinfo{person}{Rodrygo~L.T. Santos}, \bibinfo{person}{Craig
  Macdonald}, {and} \bibinfo{person}{Iadh Ounis}.}
  \bibinfo{year}{2010}\natexlab{}.
\newblock \showarticletitle{Exploiting Query Reformulations for Web Search
  Result Diversification}. In \bibinfo{booktitle}{\emph{Proceedings of the 19th
  International Conference on World Wide Web}} (Raleigh, North Carolina, USA)
  \emph{(\bibinfo{series}{WWW '10})}. \bibinfo{publisher}{Association for
  Computing Machinery}, \bibinfo{address}{New York, NY, USA},
  \bibinfo{pages}{881–890}.
\newblock
\showISBNx{9781605587998}
\urldef\tempurl%
\url{https://doi.org/10.1145/1772690.1772780}
\showDOI{\tempurl}


\bibitem[\protect\citeauthoryear{Sutskever, Vinyals, and Le}{Sutskever
  et~al\mbox{.}}{2014}]%
        {seq2seq}
\bibfield{author}{\bibinfo{person}{Ilya Sutskever}, \bibinfo{person}{Oriol
  Vinyals}, {and} \bibinfo{person}{Quoc~V. Le}.}
  \bibinfo{year}{2014}\natexlab{}.
\newblock \showarticletitle{Sequence to Sequence Learning with Neural
  Networks}. In \bibinfo{booktitle}{\emph{Proceedings of the 27th International
  Conference on Neural Information Processing Systems - Volume 2}} (Montreal,
  Canada) \emph{(\bibinfo{series}{NIPS'14})}. \bibinfo{publisher}{MIT Press},
  \bibinfo{address}{Cambridge, MA, USA}, \bibinfo{pages}{3104–3112}.
\newblock


\bibitem[\protect\citeauthoryear{Tandon, Indrakanti, Jaiswal, Strunjas, and
  Kannadasan}{Tandon et~al\mbox{.}}{2020}]%
        {tandon2020addressing}
\bibfield{author}{\bibinfo{person}{Shubhangi Tandon},
  \bibinfo{person}{Saratchandra Indrakanti}, \bibinfo{person}{Amit Jaiswal},
  \bibinfo{person}{Svetlana Strunjas}, {and}
  \bibinfo{person}{Manojkumar~Rangasamy Kannadasan}.}
  \bibinfo{year}{2020}\natexlab{}.
\newblock \showarticletitle{Addressing Purchase-Impression Gap through a
  Sequential Re-ranker}.
\newblock \bibinfo{journal}{\emph{arXiv preprint arXiv:2010.14570}}
  (\bibinfo{year}{2020}).
\newblock


\bibitem[\protect\citeauthoryear{Wang, Yin, Jie, Wang, Yamada, Chang, and
  Mei}{Wang et~al\mbox{.}}{2016}]%
        {wang2016beyond}
\bibfield{author}{\bibinfo{person}{Yue Wang}, \bibinfo{person}{Dawei Yin},
  \bibinfo{person}{Luo Jie}, \bibinfo{person}{Pengyuan Wang},
  \bibinfo{person}{Makoto Yamada}, \bibinfo{person}{Yi Chang}, {and}
  \bibinfo{person}{Qiaozhu Mei}.} \bibinfo{year}{2016}\natexlab{}.
\newblock \showarticletitle{Beyond ranking: Optimizing whole-page
  presentation}. In \bibinfo{booktitle}{\emph{Proceedings of the Ninth ACM
  International Conference on Web Search and Data Mining}}.
  \bibinfo{pages}{103--112}.
\newblock


\bibitem[\protect\citeauthoryear{Wu, Burges, Svore, and Gao}{Wu
  et~al\mbox{.}}{2010}]%
        {LambdaMART}
\bibfield{author}{\bibinfo{person}{Qiang Wu}, \bibinfo{person}{Christopher~JC
  Burges}, \bibinfo{person}{Krysta~M Svore}, {and} \bibinfo{person}{Jianfeng
  Gao}.} \bibinfo{year}{2010}\natexlab{}.
\newblock \showarticletitle{Adapting boosting for information retrieval
  measures}.
\newblock \bibinfo{journal}{\emph{Information Retrieval}} \bibinfo{volume}{13},
  \bibinfo{number}{3} (\bibinfo{year}{2010}), \bibinfo{pages}{254--270}.
\newblock


\bibitem[\protect\citeauthoryear{Xia, Xu, Lan, Guo, and Cheng}{Xia
  et~al\mbox{.}}{2016}]%
        {diverse_loss_2}
\bibfield{author}{\bibinfo{person}{Long Xia}, \bibinfo{person}{Jun Xu},
  \bibinfo{person}{Yanyan Lan}, \bibinfo{person}{Jiafeng Guo}, {and}
  \bibinfo{person}{Xueqi Cheng}.} \bibinfo{year}{2016}\natexlab{}.
\newblock \showarticletitle{Modeling document novelty with neural tensor
  network for search result diversification}. In
  \bibinfo{booktitle}{\emph{Proceedings of the 39th International ACM SIGIR
  conference on Research and Development in Information Retrieval}}.
  \bibinfo{pages}{395--404}.
\newblock


\bibitem[\protect\citeauthoryear{Xu and Li}{Xu and Li}{2007}]%
        {AdaRank}
\bibfield{author}{\bibinfo{person}{Jun Xu} {and} \bibinfo{person}{Hang Li}.}
  \bibinfo{year}{2007}\natexlab{}.
\newblock \showarticletitle{Adarank: a boosting algorithm for information
  retrieval}. In \bibinfo{booktitle}{\emph{Proceedings of the 30th annual
  international ACM SIGIR conference on Research and development in information
  retrieval}}. \bibinfo{pages}{391--398}.
\newblock


\bibitem[\protect\citeauthoryear{Zhuang, Tang, Ding, Tatikonda, Dvornek,
  Papademetris, and Duncan}{Zhuang et~al\mbox{.}}{2020}]%
        {adabelief}
\bibfield{author}{\bibinfo{person}{Juntang Zhuang}, \bibinfo{person}{Tommy
  Tang}, \bibinfo{person}{Yifan Ding}, \bibinfo{person}{Sekhar Tatikonda},
  \bibinfo{person}{Nicha Dvornek}, \bibinfo{person}{Xenophon Papademetris},
  {and} \bibinfo{person}{James~S. Duncan}.} \bibinfo{year}{2020}\natexlab{}.
\newblock \bibinfo{title}{AdaBelief Optimizer: Adapting Stepsizes by the Belief
  in Observed Gradients}.
\newblock
\newblock
\showeprint[arxiv]{2010.07468}~[cs.LG]


\end{thebibliography}
